\documentclass{article}
\usepackage{ijcai21}

\usepackage{times}

\usepackage{soul}
\usepackage{url}
\usepackage[utf8]{inputenc}
\usepackage[small]{caption}
\usepackage[colorlinks,
linkcolor=blue,
anchorcolor=blue,
citecolor=blue,
urlcolor=black,
]{hyperref}
\usepackage{graphicx}
\usepackage{amsmath}
\usepackage{booktabs}
\usepackage{amssymb}
\usepackage{multirow}
\title{SGEN: Single-cell Sequencing Graph Self-supervised Embedding Network}

\author{
Ziyi Liu$^1$\and
Minghui Liao$^1$\and
Fulin luo$^{2}$\footnote{Contact Author}\And
Bo Du$^1$\footnote{Contact Author}\\
\affiliations
$^1$National Engineering Research Center for Multimedia Software, Institute of Artificial Intelligence, School of Computer Science, and Hubei Key Laboratory of Multimedia and Network Communication Engineering, Wuhan University, Wuhan, 430072, China\\
$^2$State Key Laboratory of Information Engineering in Surveying, Mapping, and Remote Sensing, Wuhan University, Wuhan 430079, China\\
\emails
}

\begin{document}

\maketitle

\begin{abstract}
Single-cell sequencing has significant role to explore biological processes such as embryonic development, cancer evolution and cell differentiation. These biological properties can be presented by two-dimensional scatter plot. However, single-cell sequencing data generally has very high dimensionality. Therefore, dimensionality reduction should be used to process the high dimensional sequencing data for 2D visualization and subsequent biological analysis. The traditional dimensionality reduction methods, which don’t consider the structure characteristics of single-cell sequencing data, are difficult to reveal the data structure in the 2D representation. In this paper, we develop a 2D feature representation method based on graph convolutional networks (GCN) for the visualization of single-cell data, termed single-cell sequencing graph embedding networks (SGEN). This method constructs the graph by the similarity relationship between cells and adopts GCN to analyze the neighbor embedding information of samples, which makes the similar cell closer to each other on the 2D scatter plot. The results show SGEN achieves obvious 2D distribution and preserves the high-dimensional relationship of different cells. Meanwhile, similar cell clusters have spatial continuity rather than relying heavily on random initialization, which can reflect the trajectory of cell development in this scatter plot.
\end{abstract}

\section{Introduction}

Complex biological tissues are composed of functionally diverse, heterogeneous populations of cells. Single-cell sequencing\cite{2016Single}, which gives all the transcriptome or genome information of individual cell rather than bulk samples, provides cell-specific insights, including monitoring abnormal cells\cite{franke2006reconstruction}, tracking cell development \cite{hubert1985comparing}, and detecting cell responses to environmental disturbances \cite{William1971Objective}. It is now widely used in many biological fields to analyze the biological status of individual cells, including cancer biology \cite{wang2014clonal}, immunology \cite{Stubbington2017Single}, and metagenomics \cite{2011Single}. Single-cell sequencing can also identify cells with different gene expression in different environments to explore the causes of differential expression \cite{Dominic0Single}.  	

In single-cell sequencing data, each cell is described as a vector consisting of the expression values of all genes. The data from single-cell sequencing are large in volume and high-dimensional, as each cell contains tens of thousands of gene expression values and the number of cells of each batch amounts to hundreds of thousands. In the subsequent data analysis, it is very time-consuming and inconvenient to operate such a huge matrix mathematically, and the redundancy caused by many strongly correlated genes results in the waste of computing resources, so an appropriate algorithm should be constructed to reduce the dimensions of the single-cell sequencing data. The other purpose of dimensionality reduction is visualization with a 2D scatter plot, where cells in similar biological status presented on the 2D scatter plot are closer to each other than those in diverse biological status. The high volume and high dimension of the single-cell sequencing data pose challenges to existing dimensionality reduction algorithms.

Linear dimensionality reduction algorithms, such as the principal component analysis (PCA) \cite{moon2017phate}, do not work well in 2 or 3 dimensions on capturing the original structural information from high dimension for the inherent non-linearity of single-cell sequencing data. Because of the high efficiency of PCA, it is often used as a pre-processing step of downstream analysis to reduce the dimensions of data to hundreds \cite{tung2017batch}. The challenge in dimensionality reduction for single-cell sequencing data lies in the preservation of the data global structure, which contains great biological significance. The better global structure makes the distribution of clusters on the 2D scatter plot obey the similarity of the biological characteristics of cell types, whereby the 2D scatter plot should be reliable. Unfortunately, so far, even nonlinear dimensionality reduction methods don't consider the global structures of single-cell data. Here we propose SGEN, a single-cell data dimensionality reduction neural network based on graph convolutional networks (GCN). We construct the graph of cells by the similarity relationship between cells and adopt GCN to get the node aggregated embedding. Aggregated region grants GCN the ability to analyze the neighbor embedding information of samples, which makes the similar cell closer to each other on the 2D scatter plot.

Compared with existing dimensionality reduction methods used in this domain, we highlight three contributions:
\begin{itemize}
	\item Our algorithm preserves the global structure of the data, which means the distance between cell clusters reflect the similarity of cell types.
	\item SGEN can be used as a parametric model to directly position new samples on the trained 2D scatter plot, whereby we can quickly detect the biological state of the new samples.
	\item SGEN can provide the fake labels of 2-dimensional cell embedding, which explicitly present the biological state of the cells, namely, the points closer to the centroid verge to the common biological status.
\end{itemize}

\section{Related works}

In recent years, dimensional reduction of single-cell sequencing data for visualization has been a popular research topic. Several traditional machine learning methods have been proposed to visualize the 2d distribution of cells for better biological analysis. Meanwhile, deep learning methods haven’t been widely used for the visualization target.

t-distributed stochastic neighborhood embedding (t-SNE) \cite{Laurens2008Visualizing} is currently the most commonly used technology in single-cell analysis.
t-SNE has strong ability in preserving microstructure, but it is at the expense of its ability to preserve global structure \cite{Wattenberg2016How}. In other words, t-SNE can put similar cells close together on a 2D scatter plot, but it cannot put diverse cells far apart, which will make the position relations between cell clusters in the t-SNE plot unreliable \cite{Wattenberg2016How}.
\cite{2019The} proposed three improvements for the traditional t-SNE to alleviate this problem, including PCA initialisation, a high learning rate, and multi-scale similarity kernels. The heavily optimized fourier-interpolated t-SNE (FIt-SNE) proposed by \cite{2017Efficient} is widely used in single-cell data analysis because it greatly reduces the run time of t-SNE.
PHATE \cite{moon2017phate}, using the manifold distance to measure the difference of samples, can encodes the relevant information with fewer dimensions via multidimensional scaling (MDS). With diffusion map, PHATE works well in tracking trends of the data \cite{2016Diffusion}. \cite{Etienne2018Dimensionality} proposed UMAP, a dimensionality reduction model which greatly reduces running time, but it does not effectively separate clusters of cells, which means the distance between clusters cannot reflects the difference between cell types. scvis \cite{2018Interpretable}, as deep generative models, can preserve the global structure of data and greatly extract interpretation of projected structures, but the algorithm is time-consuming, and even the efficiency is not as good as t-SNE on small data sets.

In recent years, graph neural network(GNN) has been developed to learn the topological information of data, which is widely used in the fields of social sciences\cite{kipf2016semi}, knowledge graphs\cite{schlichtkrull2018modeling}, chemistry\cite{duvenaud2015convolutional}.
In the case of fixed-size graphs, a series of convolutional neural networks based on the spectral representation of the graphs have been applied on the node classification.
Specifically, Kipf \& Welling\cite{kipf2016semi} proposed a simplified spectral neural network using 1-hop filters to address overfitting problem and minimize the number of operations.

\section{Network architecture}
\begin{figure*}[t]
	\centering	
	\includegraphics[width=\linewidth]{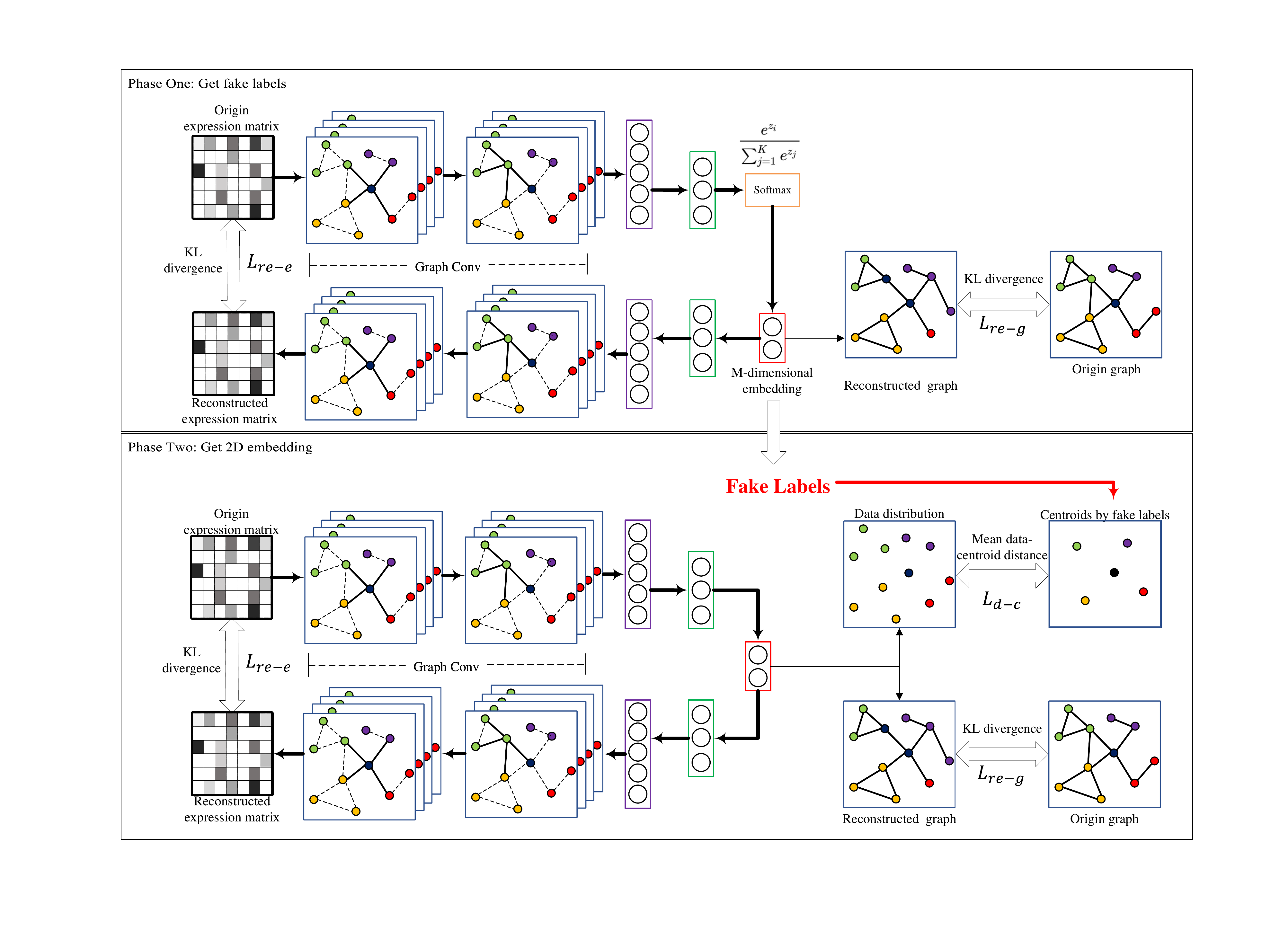}
	\caption{Network architecture of SGEN. The origin gene expression is inputed into network for reconstructed expression output. Different colored nodes in the graph represent cells of different categories. We get the low-dimensional embedding vector from the output of the encoder. The embedding vector are then used to compute the data-centroid loss and graph reconstruction loss. }
	\label{fig:Network architecture}
\end{figure*}
SGEN is an encoder-decoder model(Figure \ref{fig:Network architecture}). We aim to get a low-dimensional embedding that preserves enough information to construct the original gene expression. We construct the cells similarity graph for graph convolutional operation. Then, the topological aggregated features outputted by graph convolutional operation are processed via a fully connected auto-encoder.
We construct the fake similarity graph and compute the Kullback-Leibler divergence between the true and fake graphs to get the graph construction loss.
Besides, we get the fake cluster assignment of cells and compute the mean distance of data-centroid distance to get the data-centroid loss. Lastly, we use backpropagation to train single-cell sequencing graph embedding networks according to the expression reconstruction loss, graph construction loss and data-centroid loss.

\section{Methodology}
\subsection{Feature selection}
In our study, not all gene expression contributes to the single cell embedding cluster. We don’t aim to focus on the zero expression genes. Any gene that has high dropout rate and high non-zero mean expression could potentially be a marker of some particular subpopulation. To analyze potential contribution of each gene $g$, we compute the fraction of zero counts
\begin{equation}d_{g}=\frac{1}{n} \Sigma_{i} I\left(X_{i g}=0\right)\end{equation}
and the mean log non-zero expression
\begin{equation}m_{g}=\left\{\log _{2} X_{i g} \mid X_{i g}>0\right\}\end{equation}
where $I$ is the counting function.

In order to select the genes that have special value for clustering. We intend to find the gene that have high mean log non-zero expression and high zero counts. However, there is a strong negative relationship between $d_g$ and $m_g$ across all the remaining genes. We use the formula to get the M-dimensional features for neural networks inputs:
\begin{equation}d_{g}>\exp \left[-\left(m_{g}-b\right)\right]+0.02\end{equation}
where \emph{M} is a hyperparameter to select a pre-specified number of contributive genes, \emph{b} is the value we need to find for matching \emph{M}. This process can be done through a binary search.

\subsection{Graph construction}
To reflect the local structure in the high-dimensional space, we define the relationship of point $i$ to $j$ according to the notion of directional similarity introduced by SNE \cite{DBLP:conf/nips/HintonR02}. We construct the graph to measure similarity between each point. The node features of graph are defined by the normalization of the pre-specified number of contributive genes. We use the directional similarity to construct edges of the similarity graph.
\begin{equation}p_{i \mid j}=\frac{\exp \left(-|| x_{i}-x_{j}||^{2} / 2 \sigma_{i}^{2}\right)}{\Sigma_{k \neq i} \exp \left(-|| x_{i}-x_{k}||^{2} / 2 \sigma_{i}^{2}\right)}\end{equation}

The variance $\sigma_i^2$ of Gaussian kernel, which is often set to 30, is chosen with the criterion that the perplexity of this probability distribution equals to a pre-defined hyperparameter. The perplexity is defined as:
\begin{equation}\mathcal{P}_{i}=\exp \left(-\log (2) \sum_{j \neq i} p_{j \mid i} \log _{2} p_{j \mid i}\right)\end{equation}

Because we only focus on the similarity of different points, we set $p_{i|i}$ as 0. For the computational convenience, we get the undirectional similarity A by  $A=(P+P^{T})/2$, where $P$ is the normalized directional similarity matrix, $P^T$ is the transpose of matrix $P$.

\subsection{Graph convolutional network}
For previous deep learning methods, the gene expression is inputted into the fully connected layers to get the embedding, which don’t consider the neighborhood information of points. For the GNN methods, the adjacent matrix is inputted into the GNN model to aggregate the adjacent node information. In this paper, we use GNN to learn the adjacent node information and node embedding features to analyze the similar points.

Given a graph $G=(V,E,A)$, where $V$ is a finite set of $|V|=n$ nodes, $E$ is a set of edges and $A \in \mathbb{R}^{n \times n}$ is an adjacency matrix encoding the connection weight between two nodes. For comprehension, we consider the graph convolution following general “message-passing” architecture:
\begin{equation}H^{(l)}=F\left(A, H^{(l-1)} ; \theta^{(l)}\right)\end{equation}
where $H^{(l)}\in \mathbb{R}^{n\times d} $ are the node embeddings after $l$ steps of graph convolution operation, $F$ is the graph convolution operation which is known as the message propagation function, $H^{(l-1)}$ is the output of last convolution operation, $\theta^{(l)}$ is the trainable parameters.

Many implementations of message propagation function have been proposed to get the output of the graph convolution operation. A popular method is the graph convolution network \cite{kipf2016semi}, which is implemented by linear transformations and ReLU non-linearities:
\begin{equation}H^{(l)}=\operatorname{ReLU}\left(\widetilde{D}^{-\frac{1}{2}} \tilde{A} \tilde{D}^{-\frac{1}{2}} X^{(l-1)} W^{(l)}\right)\end{equation}
where $\tilde{A}=A+I_{N}$, $\widetilde{D}_{i j}=\Sigma_{j} \tilde{A}_{i j}$ and $W^{(l)}$ is a trainable matrix. $\widetilde{D}$ is the degree matrix of $\widetilde{A}$. $\widetilde{D}^{-\frac{1}{2}} \tilde{A} \widetilde{D}^{-\frac{1}{2}}$ is a renormalization trick which is introduced to alleviate the numerical instability and exploding/vanishing gradient problem.

\subsection{Expression reconstruction loss}
The encoder processes the inputted features to get the low-dimensional embedding. If the original features and the outputted features are the same, we could say that the low-dimensional features have the ability to represent the original features. Therefore, our target is that the reconstructed expression features should be as similar to original features as possible. The reconstruction loss is defined as:
\begin{equation}\text {L}_{re-e}=K L(X, Y) \propto-\frac{1}{n} \Sigma_{i}^{n} x_{i} \log y_{i}\end{equation}
where $X$ is the original expression features, $Y$ is the reconstructed expression features. $x_{i}$ and $y_{i}$ are the $i^{t h}$ original data and reconstructed data, respectively.

\subsection{Graph reconstruction loss}
 We adopt the main idea of t-SNE, which use a t-distribution with one degree of freedom as the low-dimensional similarity kernel:
\begin{equation}q_{i j}=\frac{w_{i j}}{\Sigma_{k, k \neq i} w_{i k}}, w_{i j}=\frac{1}{1+|| z_{i}-z_{j}||^{2}}\end{equation}
where $q_{i j}$ is the similarity between the 2-dimensional embedding of the $i^{t h}$ and the $j^{t h}$ points, $z_{i}$ is the 2-dimensional embedding of the $i^{t h}$ point. Thus, we can get the similarity matrix based on the 2-dimensional embedding.

To ensure that the low-dimensional embedding remains the interactive similar relationship of the high-dimensional features, we minimize the Kullback-Leibler divergence between the low- and high- dimensional features:
\begin{equation}\operatorname{L}_{re-g}=\Sigma_{i, j} p_{i j} \log \frac{p_{i j}}{q_{i j}} \propto-\Sigma_{i, j} p_{i j} \log q_{i j}\end{equation}
where $p_{i j}$ is the similarity between the original features of the $i^{t h}$ and the $j^{t h}$ points.

\subsection{Data-centroid loss}
Although previous methods can capture and visualize the low-dimensional structures, it’s hard to distinguish local neighbor structures and the subordinate clusters.
To analyze the local property of points carefully, we train another network and save the fake labels. The networks have the same structure as SGEN and only adopt $L_{re-e}$ and $L_{re-g}$.
We input expression features and set the latent dimension as number of clusters.
We activate the latent vector using softmax activation layer. Then, we get the latent vector as fake labels for single cells. 
The detailed information is shown in supplement materials (Figure S1).

Therefore, we can get the fake centroid of cells with the each cluster by the fake labels. We aim to minimize the mean distance between the 2D embeddings and centroids.
Thus, the similar points will cluster together around the nearest centroid. Then we design a data-centroid loss function as follow:
\begin{equation}\operatorname{L}_{d-c}=\frac{1}{n} \Sigma_{i}^{n} \min _{\mathrm{j}}|| x_{i}-c_{j}||_{2}\end{equation}
where $n$ is the number of points, $x_i$ is the 2-dimensional embedding of points, $c_j$ is the 2-dimensional coordinates of the centroids computed according to the fake labels. The number of the cluster centroids is a hyperparameter, which we set to 80 in our model. To make it easier to converge, we update the centroid of cells every 5 epochs.

\subsection{Loss function}
We consider the expression reconstruction loss, graph reconstruction loss and data-centroid loss together. Thus the loss function of our method is defined as follows:
\begin{equation}
	\operatorname{Loss} = \operatorname{L}_{re-e} + \operatorname{L}_{re-g} + \lambda * \operatorname{L}_{d-c}
\end{equation}
where $\lambda$ is hyperparameter that decides the influence of the data-centroid loss, which we set to 0.1 in our paper.

\section{Experiments}
\subsection{Evaluation metric}
\begin{itemize}
\item Fisher's ratio: The ratio of intra-cluster to inter-cluster scatter matrices can be used to formulate an effective criterion of cluster relationship, which is known as Fisher’s ratio. The intra-cluster and inter-cluster scatter matrices are defined as:
\begin{equation}\mathbf{S}_{W}=\sum_{c} \sum_{i \in c}\left(\mathbf{y}_{i}-\overline{\mathbf{Y}}_{c}\right)\left(\mathbf{y}_{i}-\mathbf{Y}_{c}\right)^{T}\end{equation}
\begin{equation}\mathbf{S}_{B}=\sum_{c}\left(\mathbf{Y}_{c}-\mathbf{Y}\right)\left(\mathbf{Y}_{c}-\mathbf{Y}\right)^{T}\end{equation}
where $\bar{Y}_{c}$ is the class-oriented average value, and $\bar{Y}$ is the overall mean of the data. The ratio determining the class separability is defined as:
\begin{equation}J_{B / W}=\operatorname{tr}\left(\mathbf{S}_{W}^{-1} \mathbf{S}_{B}\right)=\operatorname{tr}\left(\mathbf{S}_{B} \mathbf{S}_{W}^{-1}\right)\end{equation}

The larger $J_{B/W}$, the more compact points of the same clusters and the more sperate points of different clusters. We use the ratio to reveal the distribution of low-dimensional points.

\item KNC: The fraction of k-nearest class(KNC) means in the original data, which are preserved as k-nearest class means in the embedding. This is computed for class means only and averaged across all classes. For all datasets, we set the k as 10. KNC quantifies the preservation of the mesoscopic structure.

\item CPD: Spearman correlation between pairwise distances(CPD) in the high-dimensional space and in the embedding, which is computed with all cell pairs in the datasets among 1000 randomly chosen points. CPD quantifies preservation of the global, or macroscropic structure.

\end{itemize}

\subsection{Datasets}
In this study, we use four single-cell sequencing data for experiments.

\cite{Tasic2018shared} investigated the diversity of cell types across the adult mouse neocortex, collecting 23822 cells from two areas at distant poles of the mouse neocortex. In the dataset, 133 transcriptomic cell types are analyzed.

\cite{shekhar2016comprehensive} derived a digital expression matrix appreciably expressed genes across 27,499 cells after aligning reads, demultiplexing and counting UMIs. The dataset has identified 26 putative cell type clusters.

\cite{macosko2015highly} analyzed transcriptomes from 44808 mouse retinal cells and identified 39 transcriptionally distinct cell populations

Furthermore, we use Samusik$\_$01 dataset \cite{samusik2016automated} to explain the function of GSEN in tracking Cell development trajectory. Samusik$\_$01 dataset contain 86864 cells from 24 cell populations, which are all related to bone marrow hematopoiesis.

\subsection{Parameter setting}
We train our model for a maximum of 1000 epochs using Adam \cite{kingma2015adam}. For each epoch, we set the size of batched training data as 1024. For the learning rate parameters, we set the initial learning rate as 0.001 and the learning rate reduce factor as 0.5. The learning rate will decrease by learning rate reduce factor if the training loss does not decrease for 10 consecutive epochs. When the learning rate is equal to 1e-8, the training procedure will stop. Small change of the other parameters did not change the results much. We set the weight decay and dropout as 0.01 and 0.1 respectively. For baseline models, we set the parameters same as their original papers.

We run our experiments on a Ubuntu server with NVIDIA GTX 2080Ti GPU with memory 12 GB. The initial weights and bias use default setting in PyTorch.

\subsection{Preservation results}
\begin{table}[h]
	\centering
	\caption{Performance of our method and other baseline methods on Tasic, bipolar and retina dataset. The italic and bold font indicates the best and the second best among compared methods.}
	\begin{tabular}{@{}c|c|cccc@{}}
		\toprule
		\multicolumn{1}{c}{Datasets}& \multicolumn{1}{c}{Method} 	& \multicolumn{1}{c}{FR} 	& \multicolumn{1}{c}{KNC} 	& \multicolumn{1}{c}{CPD} 	& \\\midrule
		\multirow{4}{*}{Tasic}   	& t-SNE                      	& 1.3468                			& 0.6910                	& 0.5070                	& \\[2pt]
		& UMAP                    		& 2.3546                			& 0.6684              		& \textit{0.5133}      		& \\[2pt]
		& scvis                   		& \textit{2.4530}     				&\textit{0.7308}      		& 0.4612                	& \\[2pt]
		& SGEN                       	& \textbf{2.4688}       			& \textbf{0.7376}       	& \textbf{0.6214}       	& \\[2pt] \hline
		&           					&                					&          					&          					& \\[-6pt]
		\multirow{4}{*}{Bipolar} 	& t-SNE                      	& 0.0251                			& 0.4462               	 	& 0.5003    				& \\[2pt]
		& UMAP                       	& \textbf{0.0773}       			& 0.5115                	& 0.5881               		& \\[2pt]
		& scvis                      	& \textit{0.07538}      			& \textbf{0.6346}       	& \textbf{0.6652}      		& \\[2pt]
		& SGEN                       	& 0.05098               			& \textit{0.5150}           & \textit{0.6588}           & \\[2pt] \hline
		&           					&                					&          					&           				& \\[-6pt]
		\multirow{4}{*}{Retina}  	& t-SNE                      	& 0.0087                			& 0.7615                	& 0.7275   					& \\[2pt]
		& UMAP                       	& 0.0165                			& 0.7436                 	& 0.8982         			& \\[2pt]
		& scvis                      	& \textit{0.0194}                	& \textit{0.7769}           & \textit{0.9140}           & \\[2pt]
		& SGEN                       	& \textbf{0.0224}       			& \textbf{0.7923}       	& \textbf{0.9155}      		& \\[2pt] \bottomrule
	\end{tabular}
	\label{Tab:Metric comparison}
\end{table}
We compare our method with other baseline methods on Tasic, bipolar and retina datasets. The detailed results are listed in the Table \ref{Tab:Metric comparison}.
For the Tasic and retina datasets, GSEN performs the best on Fisher's ratio(FR), KNC and CPD.
For the bipolar dataset, SGEN gets the second best KNC and CPD.

In general, SGEN performs well in CPD on the three datasets, which means that SGEN can highly preserve the global structure of cells in the low-dimensional embedding. The relationship between different clusters can be well reflected in the 2D visualization.

Besides, SGEN gets better FR on single-cell datasets. On the Tasic and retina datasets, we get the best performance, which can be seen from Figure \ref{fig:visualization comprison} that different cells are separated apart clearly and similar cells are pulled in together. Without the cell labels, we can easily get the cell type for better biological research. The results of FR reflect that our methods can perform well on the perservation of local structure information.

Furthremore, our method gets the highest KNC across the Tasic and retina datasets. The neighbor clusters of different cell type are well preserved through SGEN. Although scvis gets the performance on the metrics, too many scattered points exist in the plot and the distribution of different cells aren't clear. When the color of the points are the same in the plot, it's hard to tell which clusters the cells should belong to.

\subsection{Loss influence}
\begin{figure}[!htbp]
	\centering	\includegraphics[width=\linewidth]{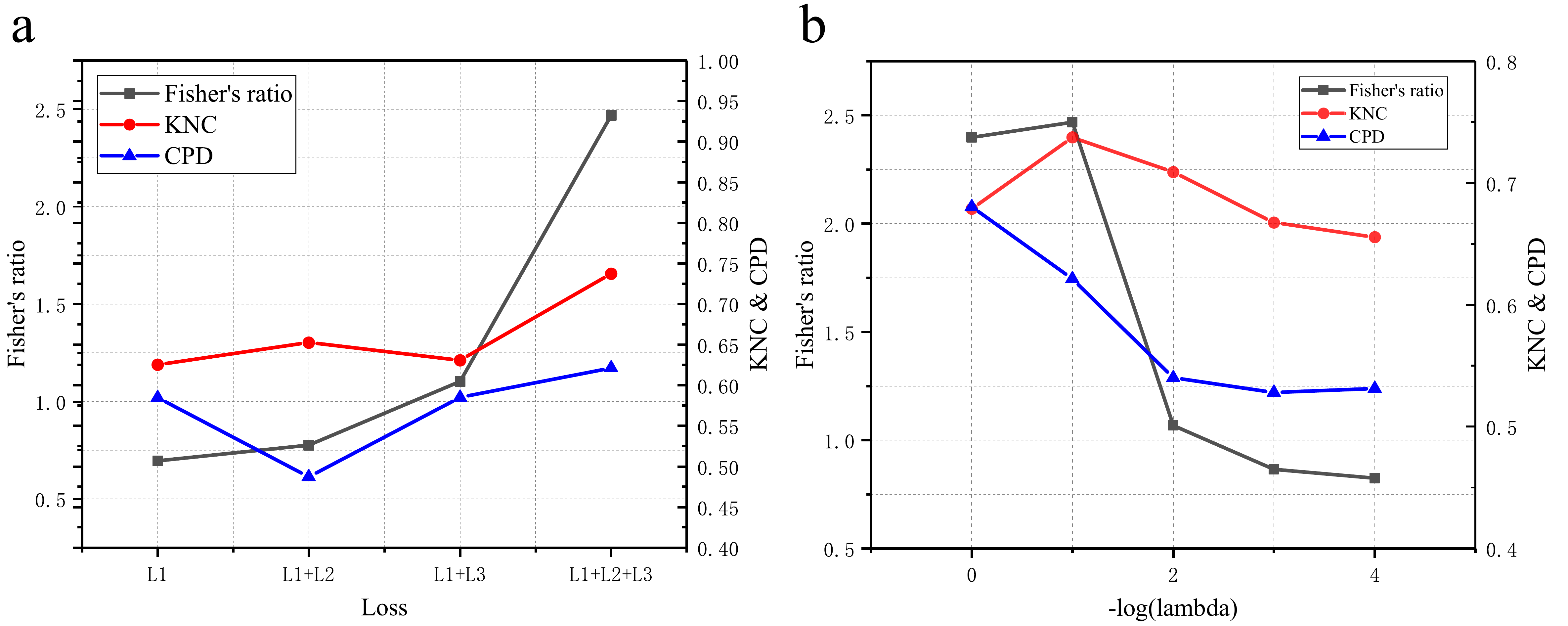}
	\caption{ The influence of loss setting on embedding quality on the Tasic dataset.  (a) Embedding quality results of different loss combination. L1, L2 and L3 represent $L_{re-e}$, $L_{re-g}$ and $L_{d-c}$, respectively. (b) Embedding quality results of different $\lambda$, where $\lambda$ decides the influence of the data-centroid loss. }
	\label{fig:parameter influence.pdf}
\end{figure}
We explain the function of each loss designed in GSEN. The detailed results can be seen in Figrue \ref{fig:parameter influence.pdf}.

From Figure \ref{fig:parameter influence.pdf}a, Fisher's ratio increases a lot when $L_{re-g}$ and $L_{d-c}$ are plused to constraint the converge of the network. Each of $L_{re-g}$ and $L_{d-c}$ can benefit KNC. Although $L_{re-g}$ makes GSEN perform worse at CPD, integration of $L_{re-g}$ and $L_{d-c}$ can boost the model performance. It can be demonstrated from Figrue \ref{fig:parameter influence.pdf} that $L_{re-g}$ and $L_{d-c}$ are relatively independent and have complementary effects for GSEN.

From Figure \ref{fig:parameter influence.pdf}b, $L_{d-c}$ will constraint the model to preserve more local structure of cells distribution.
Besides, $\lambda$ can benefit the presearvation of global structure.
However, when $\lambda$ is too large the KNC will begin to decrease, which means that our model will damage mesoscopic structure.
To make GSEN focus on local, mesoscopic and global structure together, we often set $\lambda$ to 0.1.

\subsection{Visualization comprison}
\begin{figure*}[!htbp]
	\centering	\includegraphics[width=0.8\linewidth]{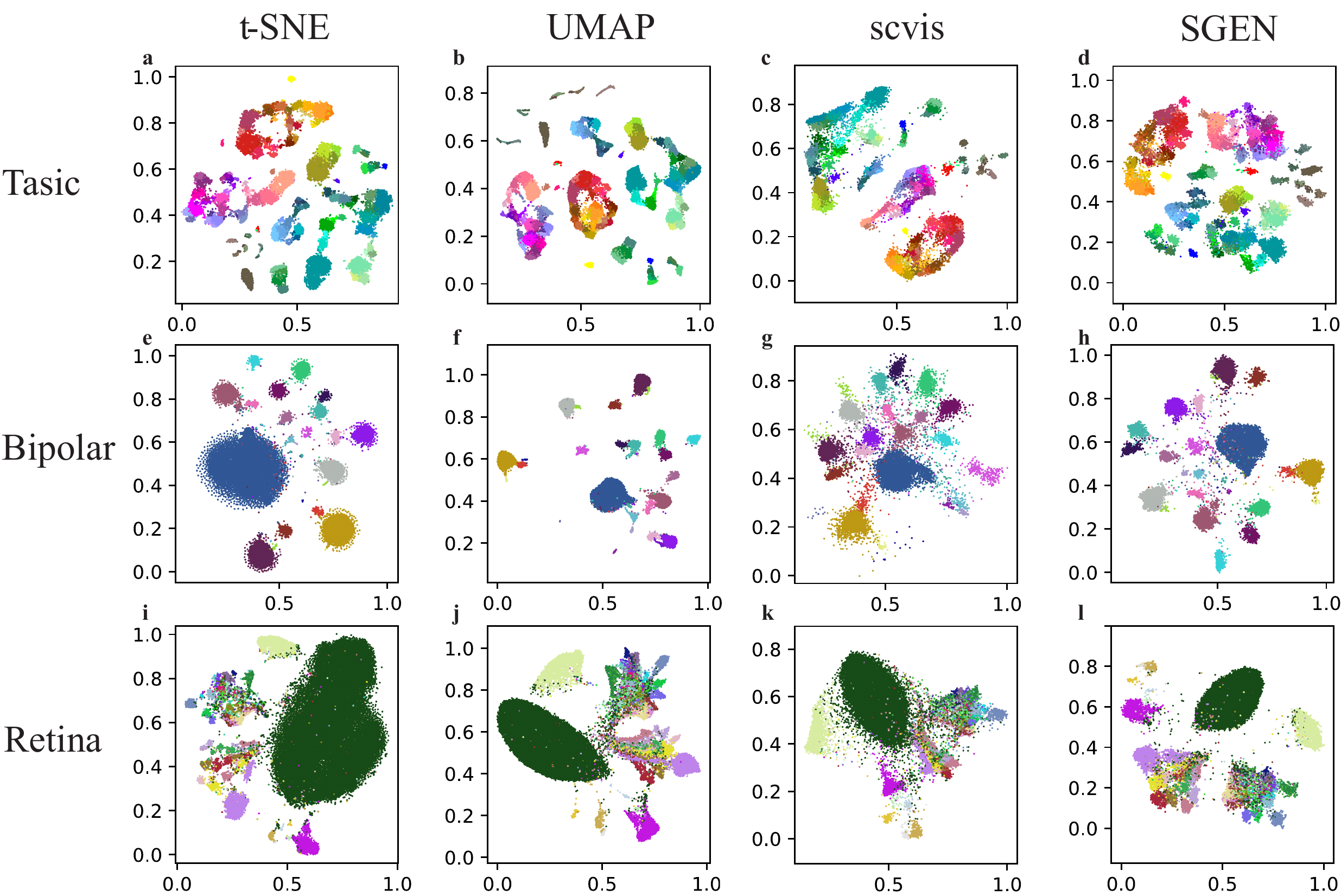}
	\caption{ 2D visualization of t-SNE, UMAP, scvis and SGEN on Tasic, bipolar and retina datasets, respectively. The color of different cells in the Tasic dataset are set according to the similarity, which means similar cell types share similar colors. The color of clusters in the bipolar and retina datasets are randomlized. All the 2D embedding points are normalized between 0 and 1 for better comprison. }
	\label{fig:visualization comprison}
\end{figure*}

\begin{figure}[t]
	\centering	\includegraphics[width=0.75\linewidth]{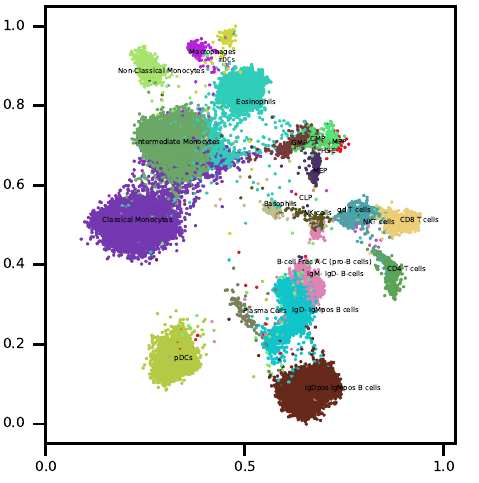}
	\caption{ SGNE embeddings of Samusik$\_$01 dataset, the color of clusters is randomlized. The cell types are annotated at the centroids of corresponding clusters. }
	\label{fig:Biological analysis.pdf}
\end{figure}
In the Tasic data, there are three main groups of cells, excitatory neurons (cold colours), inhibitory neurons (warm colours), and non-neural cells such as astrocytes or microglia (grey/brown colours).
The three groups have subclusters, which are again composed of several similar cell types. This hierarchy can be clearly seen from SGEN (Figure \ref{fig:visualization comprison}d), but it is almost invisible from UMAP (Figure \ref{fig:visualization comprison}b).
t-SNE (Figure \ref{fig:visualization comprison}a) and scvis(Figure \ref{fig:visualization comprison}c) can roughly indicate the three types of excitatory neurons, inhibitory neurons, and non-neural cells, but the distance between subclusters and between cell clusters within a subclusters cannot reflect the difference of cell types, misrepresenting the hierarchy of cell types.
For Bipolar and Retina dataset, scvis has so many out-of-cluster points that cannot be placed in the corresponding cluster(Figure \ref{fig:visualization comprison}g, k).
Although UMAP can tightly cluster the cell in common types, as in the Tasic dataset, it does not present the global structure of the data (Figure \ref{fig:visualization comprison}f, j), and the position relation between clusters is dependent on the random initial condition.
Both SGEN and t-SNE (the current mainstream applications in this field) have performed well(Figure \ref{fig:visualization comprison}e, h, i, l).

\subsection{Biological analysis}
In order to further verify that SGEN can make the distance of point clusters on the 2D scatter plot obey the difference of cell types, we used SGEN to reduce the dimension to 2-dimension on the Samusik$\_$01 dataset(Figure \ref{fig:Biological analysis.pdf}). To our surprise, we found that the position of corresponding cell clusters on the 2D scatter plot can form the developmental trajectory of cells. First of all, it is obvious that the subsets belonging to B cells and T cells are all close together without overlapping. Next, a differentiation trajectory was observed. Hematopoietic stem cells (HSCs) and multipotent progenitors (MPPs) as stem cells have similar expression characteristics and exist partial overlap. HSCs can differentiate into common myeloid progenitors (CMPs). CMPs then led to Granulocyte -myeloid progenitors (GMPs), which has two directions of differentiation, classical monocytes and intermediate monocytes, and continuous distribution points between classical monocytes and CMPs may be the cells in the process of differentiation. The above cell differentiation processes are shown in a continuous track on the 2D scatter plot. Moreover, we can further confirm that SGEN makes the dissimilar cells far away while the similar cell close to each other on the 2D scatter plot , for example, nonclassical monocytes and PDCs are far away from the surrounding clusters , respectively.

\section{Conclusion}
In this study, we develop a novel deep learning model based GCN to get the 2D embedding for the visualization of single-cell data.
We construct the graph by the similarity relationship between cells and adopt GCN to analyze the neighbor embedding information of samples, which makes the similar cell closer to each other on the 2D scatter plot.
The results show SGEN achieves obvious 2D distribution and preserves the high-dimensional relationship of different cells.
Furthermore, we use 2D embedding of GSEN for biological research and get the cell information that is consistent with the biological knowledge from existing literatures.

\clearpage
\small
\bibliographystyle{named}
\bibliography{ijcai21}

\begin{thebibliography}{}

\bibitem[\protect\citeauthoryear{Abelson \bgroup \em et al.\egroup
  }{1985}]{abelson-et-al:scheme}
Harold Abelson, Gerald~Jay Sussman, and Julie Sussman.
\newblock {\em Structure and Interpretation of Computer Programs}.
\newblock MIT Press, Cambridge, Massachusetts, 1985.

\bibitem[\protect\citeauthoryear{Baumgartner \bgroup \em et al.\egroup
  }{2001}]{bgf:Lixto}
Robert Baumgartner, Georg Gottlob, and Sergio Flesca.
\newblock Visual information extraction with {Lixto}.
\newblock In {\em Proceedings of the 27th International Conference on Very
  Large Databases}, pages 119--128, Rome, Italy, September 2001. Morgan
  Kaufmann.

\bibitem[\protect\citeauthoryear{Brachman and
  Schmolze}{1985}]{brachman-schmolze:kl-one}
Ronald~J. Brachman and James~G. Schmolze.
\newblock An overview of the {KL-ONE} knowledge representation system.
\newblock {\em Cognitive Science}, 9(2):171--216, April--June 1985.

\bibitem[\protect\citeauthoryear{Gottlob \bgroup \em et al.\egroup
  }{2002}]{gls:hypertrees}
Georg Gottlob, Nicola Leone, and Francesco Scarcello.
\newblock Hypertree decompositions and tractable queries.
\newblock {\em Journal of Computer and System Sciences}, 64(3):579--627, May
  2002.

\bibitem[\protect\citeauthoryear{Gottlob}{1992}]{gottlob:nonmon}
Georg Gottlob.
\newblock Complexity results for nonmonotonic logics.
\newblock {\em Journal of Logic and Computation}, 2(3):397--425, June 1992.

\bibitem[\protect\citeauthoryear{Levesque}{1984a}]{levesque:functional-foundations}
Hector~J. Levesque.
\newblock Foundations of a functional approach to knowledge representation.
\newblock {\em Artificial Intelligence}, 23(2):155--212, July 1984.

\bibitem[\protect\citeauthoryear{Levesque}{1984b}]{levesque:belief}
Hector~J. Levesque.
\newblock A logic of implicit and explicit belief.
\newblock In {\em Proceedings of the Fourth National Conference on Artificial
  Intelligence}, pages 198--202, Austin, Texas, August 1984. American
  Association for Artificial Intelligence.

\bibitem[\protect\citeauthoryear{Nebel}{2000}]{nebel:jair-2000}
Bernhard Nebel.
\newblock On the compilability and expressive power of propositional planning
  formalisms.
\newblock {\em Journal of Artificial Intelligence Research}, 12:271--315, 2000.

\end{thebibliography}


\begin{thebibliography}{}

\bibitem[\protect\citeauthoryear{Ding \bgroup \em et al.\egroup
  }{2018}]{2018Interpretable}
Jiarui Ding, Anne Condon, and Sohrab~P. Shah.
\newblock Interpretable dimensionality reduction of single cell transcriptome
  data with deep generative models.
\newblock {\em Nature Communications}, 9(1):2002, 2018.

\bibitem[\protect\citeauthoryear{Duvenaud \bgroup \em et al.\egroup
  }{2015}]{duvenaud2015convolutional}
David~K Duvenaud, Dougal Maclaurin, Jorge Iparraguirre, Rafael Bombarell,
  Timothy Hirzel, Al{\'a}n Aspuru-Guzik, and Ryan~P Adams.
\newblock Convolutional networks on graphs for learning molecular fingerprints.
\newblock In {\em Advances in neural information processing systems}, pages
  2224--2232, 2015.

\bibitem[\protect\citeauthoryear{Etienne \bgroup \em et al.\egroup
  }{2018}]{Etienne2018Dimensionality}
Etienne, Becht, Leland, McInnes, John, Healy, Charles-Antoine, Dutertre,
  Immanuel, and W~and.
\newblock Dimensionality reduction for visualizing single-cell data using umap.
\newblock {\em Nature Biotechnology}, 2018.

\bibitem[\protect\citeauthoryear{Franke \bgroup \em et al.\egroup
  }{2006}]{franke2006reconstruction}
Lude Franke, Harm Van~Bakel, Like Fokkens, Edwin~D De~Jong, Michael
  Egmont-Petersen, and Cisca Wijmenga.
\newblock Reconstruction of a functional human gene network, with an
  application for prioritizing positional candidate genes.
\newblock {\em The American Journal of Human Genetics}, 78(6):1011--1025, 2006.

\bibitem[\protect\citeauthoryear{Gawad \bgroup \em et al.\egroup
  }{2016}]{2016Single}
Charles Gawad, Winston Koh, and Stephen~R. Quake.
\newblock Single-cell genome sequencing: current state of the science.
\newblock {\em Nature Reviews Genetics}, 2016.

\bibitem[\protect\citeauthoryear{Grün \bgroup \em et al.\egroup
  }{}]{Dominic0Single}
Dominic Grün, Anna Lyubimova, Lennart Kester, Kay Wiebrands, Onur Basak, Nobuo
  Sasaki, Hans Clevers, and Alexander~Van Oudenaarden.
\newblock Single-cell messenger rna sequencing reveals rare intestinal cell
  types.
\newblock {\em Nature}.

\bibitem[\protect\citeauthoryear{Haghverdi \bgroup \em et al.\egroup
  }{2016}]{2016Diffusion}
Laleh Haghverdi, Maren Büttner, F~Alexander Wolf, Florian Buettner, and
  Fabian~J Theis.
\newblock Diffusion pseudotime robustly reconstructs lineage branching.
\newblock {\em Nature Methods}, 2016.

\bibitem[\protect\citeauthoryear{Hinton and
  Roweis}{2002}]{DBLP:conf/nips/HintonR02}
Geoffrey~E. Hinton and Sam~T. Roweis.
\newblock Stochastic neighbor embedding.
\newblock In Suzanna Becker, Sebastian Thrun, and Klaus Obermayer, editors,
  {\em Advances in Neural Information Processing Systems 15 [Neural Information
  Processing Systems, {NIPS} 2002, December 9-14, 2002, Vancouver, British
  Columbia, Canada]}, pages 833--840. {MIT} Press, 2002.

\bibitem[\protect\citeauthoryear{Hubert and Arabie}{1985}]{hubert1985comparing}
Lawrence Hubert and Phipps Arabie.
\newblock Comparing partitions.
\newblock {\em Journal of classification}, 2(1):193--218, 1985.

\bibitem[\protect\citeauthoryear{Kingma and Ba}{2015}]{kingma2015adam}
Diederik~P Kingma and Jimmy~Lei Ba.
\newblock Adam: A method for stochastic gradient descent.
\newblock In {\em ICLR: International Conference on Learning Representations},
  2015.

\bibitem[\protect\citeauthoryear{Kipf and Welling}{2017}]{kipf2016semi}
Thomas~N. Kipf and Max Welling.
\newblock Semi-supervised classification with graph convolutional networks.
\newblock In {\em 5th International Conference on Learning Representations,
  {ICLR} 2017, Toulon, France, April 24-26, 2017, Conference Track
  Proceedings}, 2017.

\bibitem[\protect\citeauthoryear{Kobak and Berens}{2019}]{2019The}
Dmitry Kobak and Philipp Berens.
\newblock The art of using t-sne for single-cell transcriptomics.
\newblock {\em Nature Communications}, 10(1), 2019.

\bibitem[\protect\citeauthoryear{Laurens and
  Hinton}{2008}]{Laurens2008Visualizing}
Van Der~Maaten Laurens and Geoffrey Hinton.
\newblock Visualizing data using t-sne.
\newblock {\em Journal of Machine Learning Research}, 9(2605):2579--2605, 2008.

\bibitem[\protect\citeauthoryear{Linderman \bgroup \em et al.\egroup
  }{2017}]{2017Efficient}
George~C Linderman, Manas Rachh, Jeremy~G Hoskins, Stefan Steinerberger, and
  Yuval Kluger.
\newblock Efficient algorithms for t-distributed stochastic neighborhood
  embedding.
\newblock 2017.

\bibitem[\protect\citeauthoryear{Macosko \bgroup \em et al.\egroup
  }{2015}]{macosko2015highly}
Evan~Z Macosko, Anindita Basu, Rahul Satija, James Nemesh, Karthik Shekhar,
  Melissa Goldman, Itay Tirosh, Allison~R Bialas, Nolan Kamitaki, Emily~M
  Martersteck, et~al.
\newblock Highly parallel genome-wide expression profiling of individual cells
  using nanoliter droplets.
\newblock {\em Cell}, 161(5):1202--1214, 2015.

\bibitem[\protect\citeauthoryear{Moon \bgroup \em et al.\egroup
  }{2017}]{moon2017phate}
Kevin~R Moon, David van Dijk, Zheng Wang, William Chen, Matthew~J Hirn,
  Ronald~R Coifman, Natalia~B Ivanova, Guy Wolf, and Smita Krishnaswamy.
\newblock Phate: a dimensionality reduction method for visualizing trajectory
  structures in high-dimensional biological data.
\newblock {\em BioRxiv}, page 120378, 2017.

\bibitem[\protect\citeauthoryear{Samusik \bgroup \em et al.\egroup
  }{2016}]{samusik2016automated}
Nikolay Samusik, Zinaida Good, Matthew~H Spitzer, Kara~L Davis, and Garry~P
  Nolan.
\newblock Automated mapping of phenotype space with single-cell data.
\newblock {\em Nature methods}, 13(6):493--496, 2016.

\bibitem[\protect\citeauthoryear{Schlichtkrull \bgroup \em et al.\egroup
  }{2018}]{schlichtkrull2018modeling}
Michael Schlichtkrull, Thomas~N Kipf, Peter Bloem, Rianne Van Den~Berg, Ivan
  Titov, and Max Welling.
\newblock Modeling relational data with graph convolutional networks.
\newblock In {\em European Semantic Web Conference}, pages 593--607. Springer,
  2018.

\bibitem[\protect\citeauthoryear{Shekhar \bgroup \em et al.\egroup
  }{2016}]{shekhar2016comprehensive}
Karthik Shekhar, Sylvain~W Lapan, Irene~E Whitney, Nicholas~M Tran, Evan~Z
  Macosko, Monika Kowalczyk, Xian Adiconis, Joshua~Z Levin, James Nemesh,
  Melissa Goldman, et~al.
\newblock Comprehensive classification of retinal bipolar neurons by
  single-cell transcriptomics.
\newblock {\em Cell}, 166(5):1308--1323, 2016.

\bibitem[\protect\citeauthoryear{Stubbington \bgroup \em et al.\egroup
  }{2017}]{Stubbington2017Single}
Michael J.~T. Stubbington, Orit Rozenblatt-Rosen, Aviv Regev, and Sarah~A.
  Teichmann.
\newblock Single-cell transcriptomics to explore the immune system in health
  and disease.
\newblock {\em Science}, 358(6359):58, 2017.

\bibitem[\protect\citeauthoryear{Tasic \bgroup \em et al.\egroup
  }{2018}]{Tasic2018shared}
Bosiljka Tasic, Zizhen Yao, Lucas~T Graybuck, Kimberly~A Smith, Thuc~Nghi
  Nguyen, Darren Bertagnolli, Jeff Goldy, Emma Garren, Michael~N Economo,
  Sarada Viswanathan, et~al.
\newblock Shared and distinct transcriptomic cell types across neocortical
  areas.
\newblock {\em Nature}, 563(7729):72--78, 2018.

\bibitem[\protect\citeauthoryear{Tung \bgroup \em et al.\egroup
  }{2017}]{tung2017batch}
Po-Yuan Tung, John~D Blischak, Chiaowen~Joyce Hsiao, David~A Knowles,
  Jonathan~E Burnett, Jonathan~K Pritchard, and Yoav Gilad.
\newblock Batch effects and the effective design of single-cell gene expression
  studies.
\newblock {\em Scientific reports}, 7:39921, 2017.

\bibitem[\protect\citeauthoryear{Wang \bgroup \em et al.\egroup
  }{2014}]{wang2014clonal}
Yong Wang, Jill Waters, Marco~L Leung, Anna Unruh, Whijae Roh, Xiuqing Shi, Ken
  Chen, Paul Scheet, Selina Vattathil, Han Liang, et~al.
\newblock Clonal evolution in breast cancer revealed by single nucleus genome
  sequencing.
\newblock {\em Nature}, 512(7513):155--160, 2014.

\bibitem[\protect\citeauthoryear{Wattenberg \bgroup \em et al.\egroup
  }{2016}]{Wattenberg2016How}
Martin Wattenberg, Fernanda Viégas, and Ian Johnson.
\newblock How to use t-sne effectively.
\newblock {\em Distill}, 1(10), 2016.

\bibitem[\protect\citeauthoryear{William \bgroup \em et al.\egroup
  }{1971}]{William1971Objective}
William, M., and Rand.
\newblock Objective criteria for the evaluation of clustering methods.
\newblock {\em Journal of the American Statistical Association},
  66(336):846--850, 1971.

\bibitem[\protect\citeauthoryear{Yoon \bgroup \em et al.\egroup
  }{2011}]{2011Single}
Hwan~Su Yoon, Dana~C. Price, Ramunas Stepanauskas, Veeran~D. Rajah, Michael~E.
  Sieracki, William~H. Wilson, Eun~Chan Yang, Siobain Duffy, and Debashish
  Bhattacharya.
\newblock Single-cell genomics reveals organismal interactions in uncultivated
  marine protists.
\newblock {\em Science}, 332(6030):714--7, 2011.

\end{thebibliography}

\end{document}